\documentclass{ws-ijmpa}

\begin{document}

\markboth{Pakvasa}
{Neutrino Flavor Detection at Neutrino Telescopes and Its Uses (Paper's Title)}

%
\catchline{}{}{}{}{}
%

\title{NEUTRINO FLAVOR DETECTION AT NEUTRINO
TELESCOPES AND ITS USES} 

\author{SANDIP PAKVASA}

\address{Department of Physics and Astronomy, University of Hawaii\\
Honolulu, HI 96822}
%
%


\maketitle


\section{Introduction}	
We assume that sources of high energy astrophysical
neutrinos exist with detectable fluxes. The existence of gamma ray
sources with energies of upto 100s of TeV provides some evidence that
this assumption may be correct\cite{HESS}, assuming that the gamma rays are coming 
from $\pi^0$'s. We also need to assume that in the not too distant future, 
large  enough detectors, well instrumented and with good 
angular and energy resolution
will be operating(hopefully of several KM3
size)\cite{icecube,auger,anita}. 
We assume also that neutrino
signals will be seen, and furthermore that the detectors will have the ability
to distinguish flavors(at the moment this is only assured for $H_2O \ $\^{c} 
detectors). If all these optimistic assumptions turn out to be valid, there are a
number of uses that these detectors can be put to\cite{pakvasa}. I would like to discuss 
some of them.

\section{Astrophysical neutrino flavor content}

Since most of the sources are tenous and emit neutrinos
from $\pi/K$ decays via the $\pi-\mu-e$ chain, we expect that the ratio of 
$\nu_e$ to $\nu_\mu$ is 1:2; furthermore, estimates of $\nu_\tau$ 
production even 
at very high energies yield very small fraction of $\nu_\tau s$\cite{learned1}. Hence,
for practical purposes the flavor ratio produced in such "conventional"
sources(either via p-p or $\gamma$-p collisions) is $\nu_e:\nu_\mu:\nu_\tau
= 1:2:0$. The  mixture of $\nu's$ and $\bar{\nu}'s$ is expected to be 1:1
with some exceptions. There are also sources in which muons lose energy 
by magnetic fields or other means(called damped muon sources); in this case the
flavor mix becomes $\nu_e:\nu_\mu:\nu_\tau = 0:1:0$, this can be an
energy dependent 
effect, making the flavor mix energy dependent\cite{rachen}. 
There may be sources, in which 
the dominant component is from neutron decays, 
resembling a "beta beam"\cite{goldberg}, with 
the resultant mix being: $\nu_e:\nu_\mu:\nu_\tau = 1:0:0.$ 
If the density is high  enough for pions to interact 
before decaying, then heavy flavors dominate
and the flavor mix becomes: $\nu_e:\nu_\mu:\nu_\tau = 1:1: 0$. 
This is the case,for
example in the so-called slow-jet supernovas\cite{razzagne}.
              
There are also the neutrinos emitted as a by product in the 
GZK\cite{greisen} reaction, properly called the 
BZ(Berezinsky-Zatsepin)\cite{venya} neutrinos. The reaction
is of course $p + \gamma \rightarrow \Delta^+\rightarrow n + \pi^+$. 
In this case the flavor mix
depends on the energy\cite{engel}. Below about 100 PeV, it is a 
pure "beta beam" with
$\nu_e:\nu_\mu:\nu_\tau = 1:0:0$; and above 100 PeV it is conventional 
$\nu_e:\nu_\mu:\nu_\tau = 1:2:0$.

\section{Effect of Oscillations}

The current knowledge of neutrino masses and mixing\cite{MNS} can
be summarized
as follows\cite{concha}. The mixing matrix elements are
given to a very good approximation by the so-called tri-bi-maximal
matrix\cite{harrison}.
The bound on the element  $|U_{e3}|$ comes from the CHOOZ experiment\cite
{chooz} and is given by $|U_{e3}| < 0.17 $.
The mass spectrum has two possibilities: normal or 
inverted. 
The mass differences are given by $|\delta m^2_{32}| \sim 
2.4.10^{-3} eV^2$(with the + sign corresponding to normal
hierarchy and - sign to the inverted one) and 
$\delta m_{21}^2 \sim +7.6.10^{-5} eV^2$.  Since $\delta m^2 
L/4E$ for 
the distances to GRB's and AGN's (even for energies up to 
and beyond PeV) 
is very large $(> 10^7)$ the oscillations have always 
averaged out 
and the conversion(or survival) probability is given by
\begin{eqnarray}
P_{\alpha \beta} &=& \sum_{i} | U_{\alpha i} \mid^2 \mid 
U_{\beta i} \mid^2 \
\end{eqnarray}
Assuming no significant matter effects en-route, it is easy 
to show that
the tri-bi-maximal mixing matrix  leads to a simple propagation matrix 
P, which, for any value of the solar mixing angle, 
converts a 
flux ratio of 
$\nu_e: \nu_\mu: \nu_\tau = 1:2:0$ 
into one of $1:1:1$.  Hence the flavor mix expected at 
arrival 
is simply an equal mixture of $\nu_e, \nu_\mu$ and 
$\nu_\tau$ as was observed long ago\cite{learned1,athar}.
Similarly, for the other cases neutrino mixing modifies the flavor mixes
as:
\begin{center}
\begin{tabular}[h!]{|l|l|c|} \hline
    \quad     & Initial  &   After Mixing  \nonumber \\
Damped Muon  &   (0:1:0)       &   (4:7:7) \nonumber \\
Beta Beam     &   (1:0:0)       &  (5:2:2)  \nonumber \\
Prompt       & (1:1:0)    &   (14:11:11) \\ \hline
\end{tabular}
\end{center}

If the universal flavor mix is confirmed by future 
observations, our current
knowledge of neutrino masses and mixing is reinforced and 
conventional
wisdom about the beam dump nature of the production 
process is confirmed as
well. However, it would be much more exciting to find 
deviations from it, and learn something new. How can this 
come about? I give below  a shopping list of  variety of ways 
in which this could come to pass, and what can be learned in each case.
  
 \section{Discriminating Flavors}
	  
We define two ratios to distinguish various flavor mixes as:
$f_e (=e/(e+\mu+\tau)$ and $R (=\mu/(e + \tau)$. 
Then we have the following for the
various possible flavor mixes expected at earth from various source types:
\begin{center}
\begin{tabular}[h!]{|l|l|l|} \hline
Source Type                &   $f_e$             &    R  
\\ \hline                         
Pionic                     &   0.33              &   0.5 \nonumber\\
Damped-muon                &   0.22              &   0.64 \nonumber \\
Beta-Beam                  &   0.55              &   0.29  \nonumber \\
Prompt                     &   0.39              &   0.44 \\ \hline
\end{tabular}
\end{center}
	   
 It has been shown that R and/or $f_e$ can be determined up to an accuracy of about
 0.08 or so in an ice-cube type detector\cite{beacom4}. Hence, pionic, damped muon and
Beta-beam type os sources can be distinguished but probably not the prompt.

There have been many suggestions that small deviations from 
the canonical flavor mixes can be used for a variety of 
purposes, such as determine small deviations from Tri-bi-maximal
mixing(i.e. measure $\theta_{13}$ and $\delta$), small mixing with sterile
neutrinos etc \cite{xing}. However, there are several reasons why this is
rather impractical. In addition to the limits on the precision
with which $f_e$ and/or R can be measured, there are inherent 
uncertainties in the source flavor mixes themselves. For
example, in the $\pi/K$ case the flavor mix is not expected to be 
exactly $\nu_e:\nu_\mu:\nu_\tau = 1:2:0$ but rather more like 
1:1.85:0\cite{lipari}. The main reason for this effect is the muon 
polarization in the $\pi-\mu$ decay which gives rise to a $\nu_\mu$
of lower energy than  $\nu_e$, with additional subtle effects 
in K-decays. Similarly, the flavor mixes in the damped muon
case and Beta-beam case are also not expected to give rise 
to the simple pure flavor mixes. These effects combined with the
8 \% uncertainty in the experimental determination of the flavor 
mix renders extremely difficult any attempt to measure small
effects in mixing\cite{rodejohann}.

\section{Deviation from Canonical Mix}
Let us consider how the conventional mix may undergo major deviations
which are detectable.
The possibility that the mass differences between neutrino 
mass  eigenstates are zero in vacuum (and become non-zero 
only in the presence
of matter) has been raised\cite{kaplan}. If this 
is true, then the final flavor mix 
should be the same as initial, namely: $1:2:0.$ However,
very recently, analysis of low energy atmospheric neutrino
data by Super-Kamiokande has ruled out a wide variety of
models for such behavior\cite{SK}.

Neutrino decay is another important possible way for the 
flavor mix to 
deviate significantly from the democratic 
mix\cite{beacom}.  We now know that neutrinos have non-zero 
masses and non-trivial mixing,
based on the evidence for neutrino mixing and 
oscillations from the data on
atmospheric, solar and reactor neutrinos.

Once neutrinos have masses and mixing, then in general, the heavier neutrinos 
are expected to 
decay into the lighter ones via flavor changing 
processes\cite{pakvasa2}.  
The only remaining questions are (a) whether the lifetimes are 
short enough to be phenomenologically interesting (or are 
they too long?) and (b) what are the dominant decay modes.
Since we are interested
in decay modes which are likely to have rates (or lead to 
lifetimes)  which
are phenomenologically interesting,  we can rule out 
several classes of decay
modes immediately. For example, the very strong 
constraints\cite{pakvasa2} on radiative decay
modes and on three body modes such as $\nu \rightarrow 
3\nu$ render them uninteresting.

The only decay modes 
which can have interestingly fast decay rates are two 
body 
modes such as $\nu_i \rightarrow \nu_j + x$  where 
$x$ is a very light or massless particle, e.g. a Majoron.
In general, the Majoron is a mixture of the Gelmini-
Roncadelli\cite{gelmini} and Chikasige-Mohapatra-
Peccei\cite{chicasige} type Majorons.  
Explicit models of this kind which can give rise to fast
neutrino decays have
been discussed\cite{valle}.
The models with $\Delta L = 2$ are unconstrained by $\mu$ and $\tau$ decays
which cannot be engendered by such couplings.
Both($\Delta L = 2$ and $\Delta L = 0$) kinds of models with couplings  
of $\nu_\mu$ and $\nu_e$ are constrained
by the limits on multi-body $\pi$, K decays,
and on $\mu-e$ universality violation in $\pi$ and K 
decays\cite{barger}, 
but these bounds allow fast neutrino decays.

Direct limits on such decay modes are rather weak.
Current bounds on such decay modes are as follows.  For 
the mass eigenstate $\nu_1$, the limit is about
\begin{equation}
\tau_1 \geq 10^5 \ sec /eV
\end{equation}
based on observation of $\bar{\nu}_e's$ from SN1987A 
\cite{hirata}
(assuming CPT invariance). For $\nu_2$, strong  limits can 
be deduced  from
the non-observation of solar anti-neutrinos in 
KamLAND\cite{eguchi}.
A more general but similar bound is obtained from 
an analysis of solar neutrino data\cite{bell1}. This 
bound is given by:
\begin{equation}
\tau_2 \geq 10^{-4} \ sec/eV
\end{equation}
For $\nu_3$,  one can derive 
a bound from the atmospheric neutrino observations of 
upcoming neutrinos\cite{barger1}:
\begin{equation}
\tau_3 \geq \ 10^{-10} \ sec/eV
\end{equation}

The strongest lifetime limit is thus too weak to eliminate 
the possibility of
astrophysical neutrino decay by a factor about $10^7 
\times (L/100$ Mpc) 
$\times (10$ TeV/E).  It was noted that the
disappearance of all states except $\nu_1$ would prepare a 
beam that could in principle be used to measure elements 
of the neutrino mixing matrix\cite{pakvasa3}, namely the ratios $|U_{e1}|^2 
: |U_{\mu 1}|^2 : |U_{\tau 1}|^2$.  
The possibility of measuring
neutrino lifetimes over long baselines was mentioned in 
Ref.\cite{weiler}, 
and some predictions for decay in four-neutrino models 
were given in 
Ref.\cite{keranen}.  The particular values and small 
uncertainties on the neutrino mixing parameters allow 
for the first time very distinctive signatures of the 
effects of 
neutrino decay on the detected flavor ratios.  
The expected increase in neutrino lifetime sensitivity 
(and corresponding 
anomalous 
neutrino couplings) by several orders of magnitude makes 
for a very
interesting test of physics beyond the Standard Model; a 
discovery would
mean physics much more exotic than neutrino mass and 
mixing alone.   
Because of its unique signature, neutrino decay  cannot be 
mimicked by either different neutrino flavor ratios at the 
source or other non-standard neutrino interactions.

A characteristic feature of decay is its strong energy 
dependence: 
$\exp (-Lm/E \tau)$, where $\tau$ is the rest-frame 
lifetime.  
For simplicity, we will consider the case that decays are always 
complete, i.e., that 
these exponential factors vanish.  
The simplest case (and the most generic expectation) is a 
normal hierarchy 
in which both $\nu_3$ and $\nu_2$ decay, leaving only the 
lightest stable eigenstate  $\nu_1$.  In this case the 
flavor ratio is\cite{pakvasa3} 
$|U_{e1}|^2:  |U_{\mu 1}|^2 : |U_{\tau 1}|^2$. 
Thus, if $|U_{e3}| = 0$ we have
\begin{equation}
\phi_{\nu e} :  \phi_{\nu_{\mu}} :  \phi_{\nu_{\tau}}
\simeq 4 : 1 : 1, 
\end{equation}
where we used the propagation matrix derived from the tri-bi-maximal mixing.  
Note that this is an extreme deviation of the flavor ratio 
from
that in the absence of decays.  It is difficult to imagine 
other mechanisms
that would lead to such a high ratio of $\nu_e$ to 
$\nu_\mu$.  In the case
of inverted hierarchy, $\nu_3$ is the lightest and hence 
stable state, and
so\cite{beacom} we have instead 
\begin{equation}
\phi_{\nu_{e}} :  \phi_{\nu_{\mu}} : \phi_{\nu _{\tau}} = 
|U_{e3}\mid ^2 : 
|U_{\mu 3} \mid^2 : |U_{\tau 3} \mid^2 = 0 : 1 : 1.
\end{equation}
If  $|U_{e3}| = 0$ and $\theta_{atm} = 45^0$, each mass 
eigenstate has equal
$\nu_\mu$ and $\nu_\tau$ components.  Therefore, decay 
cannot break 
the equality between the $\phi_{\nu_{\mu}}$ and 
$\phi_{\nu_{\tau}}$ 
fluxes and thus the $\phi_{\nu_{e}} : \phi_{\nu_\mu}$ 
ratio contains all the useful information.

When $|U_{e3}|$ is not zero, and the hierarchy is normal, it 
is possible to
obtain information on the values of $|U_{e3}|$ as well as 
the CPV phase $\delta$\cite{beacon}.  The flavor ratio 
$e/\mu$ varies from  4 to 10 (as $|U_{e3}|$ goes from 0 to 
0.2)  
for $\cos \delta =+1$ but from 4 to 2.5 for $\cos \delta =-
1$.  The ratio $\tau/\mu$ varies from 1 to 4 $(\cos \delta 
= +1)$ or 1 to 0.25 $(\cos \delta =-1)$ for the same range 
of $U_{e3}$.

If the decays are not complete and if the daughter does 
not carry the full
energy of the parent neutrino; the resulting flavor mix is 
somewhat
different but in any case it is still quite distinct from the 
simple $1:1:1$
mix\cite{beacom}. There is a very recent exhaustive study
of the various possibilities\cite{winter}.

If the path of neutrinos takes them thru regions with 
significant magnetic 
fields and the neutrino magnetic moments are large enough, 
the flavor mix can 
be affected\cite{enquist}.  The main effect of the passage 
thru magnetic field is the 
conversion of a given helicity into an equal mixture of 
both helicity states.
This is also true in passage thru random magnetic 
fields\cite{domokos}. It has been shown recently that the presence of
a magnetic field of a few(10 or more) Gauss at the source can make
the neutrinos decohere as they traverse cosmic distances\cite{farzan}.

If the neutrinos are Dirac particles, and all magnetic 
moments are comparable, 
then the effect of the spin-flip is to simply reduce the 
overall flux of all 
flavors by half, the other half becoming the sterile Dirac 
partners.
If the neutrinos are Majorana particles, 
the flavor composition remains 1 : 1 : 1 when it 
starts from 1 : 1 : 1, and the absolute flux remains 
unchanged.

Other neutrino properties can also affect the neutrino 
flavor mix and modify
it from the canonical 1 : 1 : 1. If neutrinos have 
flavor(and equivalence
principle) violating couplings to gravity(FVG); then there can be resonance
effects which make for one way transitions(analogues of 
MSW transitions)
e.g. $\nu_\mu \rightarrow \nu_\tau$ but not vice
versa\cite{minakata,barger3}. In case of FVG for example,
this can give rise to an anisotropic deviation of the 
$\nu_\mu/\nu_\tau$
ratio from 1, becoming less than 1 for events coming from 
the direction
towards the Great
Attractor, while remaining 1 in other 
directions\cite{minakata}. If such striking effects are not seen,
then the current bounds on such violations can be improved by
six to seven orders of magnitude.

Complete quantum decoherence would give rise to a flavor 
mix given
by $1:1:1$, which is identical to the case of averaged out 
oscillations
as we saw above. The distinction is that complete 
decoherence always
leads to this result; whereas averaged out oscillations 
lead to this
result only in the special case of the initial flavor mix 
being $1:2:0.$
To find evidence for decoherence, therefore, requires a 
source which
has a different flavor mix . One possible practical 
example is the ``beta'' beam  source
with an initial flavor mix of $1:0:0$. In this 
case  decoherence 
leads to the universal $1:1:1$ mix whereas the averaged 
out oscillations
lead to $2.5:1:1$\cite{hooper}. The two cases can be easily 
distinguished from each other.

Violations of Lorentz invariance and/or CPT invariance
can change the final flavor mix from the canonical
universal mix of $1: 1: 1$ significantly. With a specific
choice of the change in dispersion relation due to Lorentz
Invariance Violation, the effects can be dramatic. For example,
the final flavor mix at sufficiently high energies can become
$ 7: 2: 0$\cite{hooper}.

If each of the three neutrino mass eigenstates is actually 
a doublet 
with very small mass difference (smaller than $10^{-6} 
eV)$, 
then there are no current experiments  that could have 
detected this. 
Such a possibility was raised long ago\cite{bilenky1}. 
It turns out that the only way to detect such small mass 
differences $(10^{-12} eV^2 > \delta m^2 > 10^{-18} eV^2)$ 
is by measuring flavor mixes of the high energy neutrinos 
from cosmic sources.  Relic supernova neutrino
signals and AGN neutrinos are sensitive to mass difference 
squared down to $10^{-20} eV^2$ \cite{beacom3}.

Let $(\nu_1^+, \nu_2^+, \nu_3^+; \nu_1^-,  \nu_2^-, 
\nu_3^-)$ 
denote the six mass eigenstates where $\nu^+$ and $\nu^-$ 
are a 
nearly degenerate pair.  A 6x6 mixing matrix rotates the 
mass 
basis into the flavor basis.  
For pseudo-Dirac 
neutrinos, Kobayashi and Lim\cite{kobayashi} have given an 
elegant proof 
that the 6x6 matrix $V_{KL}$ takes the very simple form 
(to lowest order in $\delta m^2 / m^2$:
\begin{eqnarray}
V_{KL} = \left (
\begin{array}{cc}
U & 0 \\
0 & U_R 
\end{array} \right) \cdot
\left (
\begin{array}{cc}
V_1 & iV_1 \\
V_2  & -iV_2 
\end{array}\right),
\end{eqnarray}
where the $3\times 3$ matrix U is just the usual mixing 
(MNSP)matrix determined by the atmospheric and solar 
observations, the 
$3\times 3$ matrix $U_R$ is an unknown unitary matrix and 
$V_1$ and 
$V_2$ are the diagonal matrices 
$V_1 =$ diag $(1,1,1)/\sqrt{2}$, and $V_2$=diag$(e^{-i 
\phi_1}, 
e^{-i \phi_2}, e^{-i \phi_3})/\sqrt{2}$, with the $\phi_i$ 
being arbitrary phases.

The flavor ratios deviate from $1:1:1$ when
one or two of the pseudo-Dirac oscillation modes is 
accessible.  In
the ultimate limit where $L/E$ is so large that all three 
oscillating
factors have averaged to $\frac{1}{2}$, the flavor ratios 
return to $1:1:1$,
with only a net suppression of the measurable flux, by a 
factor of
$1/2$. As a bonus, if such small pseudo-Dirac mass differences 
exist, it would enable us to measure cosmological parameters such as
the red shift in neutrinos(rather than in photons)\cite{weiler,beacom3}.

\section{Experimental Flavor Identification} 

It is obvious from the above discussion that flavor 
identification is
crucial for the purpose at hand. In a water(or ice) cerenkov 
detector flavors can be 
identified as follows.  

The $\nu_\mu$ flux can be measured by the $\mu's$ produced 
by the charged 
current interactions and the resulting $\mu$ tracks in the 
detector which
are long  at these energies.  $\nu_{e}'{s}$ produce 
showers by both
CC and NC interactions.  The total rate for showers 
includes those
produced by NC interactions of $\nu_\mu's$ and 
$\nu_\tau's$ as well and those
have to be (and can be) subtracted off to get the real 
flux of $\nu_e's$.
Double-bang and lollipop events are signatures unique to 
$\nu'_\tau s$, made
possible by the fact that tau leptons decay before they 
lose a significant
fraction of their energy.  A double-bang event consists of 
a hadronic shower
initiated by a charged-current interaction of the 
$\nu_\tau$ followed by a
second energetic shower from the decay  of the
resulting $\tau$\cite{learned1}.  A lollipop event 
consists of the second of
the double-bang showers along with the reconstructed 
tau lepton track (the
first bang may be detected or not).  In principle, with a 
sufficient number of
events, a fairly good estimate of the flavor ratio $\nu_e: 
\nu_\mu: \nu_\tau$ 
can be reconstructed (with caveats about
uncertainties) as has been discussed recently.
Deviations of the flavor ratios 
from $1:1:1$ due to possible decays are so extreme that 
they should be
readily identifiable\cite{beacom4}. High energy 
neutrino telescopes,
such as Icecube\cite{karle}, will not have perfect ability 
to separately 
measure the neutrino flux in each flavor.  However, the 
situation is
salvageable. In the limit of $\nu_\mu - \nu_\tau$ symmetry 
the fluxes for $\nu_\mu$ and 
$\nu_\tau$ are always in the ratio 1 : 1, with or without 
decay. 
This is useful since the $\nu_\tau$ flux is the hardest to 
measure. 

Even when the  tau events are not all
identifiable, the relative number of shower events to
track events can
be related to the most interesting quantity for testing
decay scenarios,
i.e., the $\nu_e$ to $\nu_\mu$ ratio.  The precision of
the upcoming
experiments should be good enough to test the extreme
flavor ratios produced
by decays.  If electromagnetic and hadronic  showers can
be separated, then
the precision will be even better\cite{beacom4}.Comparing,
for example, the
standard flavor ratios of 1 : 1 : 1 to the
possible 4 : 1 : 1 (or $ 0 : 1: 1$ for inverted hierarchy)generated by
decay, the higher(lower)
electron neutrino
flux will result in a substantial increase(decrease) in the relative 
number of 
shower events.The measurement will be limited only by the 
energy resolution of the
detector and the ability to reduce the atmospheric 
neutrino background(which 
drops rapidly with energy and 
should be negligibly small at and above the PeV scale).

\section{Discussion and Conclusions}  

The flux ratios we discuss are energy-independent to the extent 
that the following assumptions are valid: (a)the ratios at production are energy-
independent, (b) all oscillations are averaged out, and (c) 
that all possible decays are complete.  In the standard 
scenario with only oscillations, the final flux ratios are 
$\phi_{\nu_{e}} :  \phi_{\nu_{\mu}} : 
 \phi_{\nu_{\tau}} = 1 : 1 : 1$.  In the cases with decay, 
we have found rather
different possible flux ratios, for example 4 : 1 : 1 in 
the normal hierarchy and 
0 : 1 : 1 in the inverted hierarchy.  These deviations 
from 1 : 1 : 1 
are so extreme that they should be readily measurable.

If we are very fortunate, we may be able 
to observe a
reasonable number of events from several sources (of known 
distance) 
and/or over a sufficient range in energy.  Then the 
resulting 
dependence of the flux ratio $(\nu_e/\nu_\mu)$ on L/E as 
it evolves from say 4 (or 0) to 1 can 
be clear evidence of decay and further can pin down the 
actual lifetime
instead of just placing a bound\cite{barenboin}.

To summarize, we suggest that if future measurements of 
the flavor mix at
earth of high energy astrophysical neutrinos find it to be
\begin{equation}
\phi_{\nu_{e}} / \phi_{\nu_{\mu}} / \phi_{\nu_{\tau}} = 
\alpha / 1 / 1 ;
\end{equation}
then
\begin{itemize}
\item[(i)] $\alpha \approx 1$ (the most boring case) 
confirms our knowledge of the
MNSP\cite{MNS} matrix and our prejudice about the 
production mechanism;
\item[(ii)] $\alpha \approx 1/2$ indicates that the source 
emits pure
$\nu_\mu's$ and the mixing is conventional;
\item[(iii)]$\alpha \approx 3$ from a unique direction, 
e.g. the Cygnus region, would be
evidence in favor of a pure $\bar{\nu}_e$ production as 
has been suggested
recently\cite{goldberg};
\item[(iv)] $\alpha > 1$ indicates that neutrinos are 
decaying with normal
hierarchy; and 
\item[(v)]$\alpha \ll 1$ would mean that neutrino decays 
are occurring with
inverted hierarchy;
\item[(vi)] Values of $\alpha$ which cover a broader range 
and 
deviation of the $\mu/\tau$ ratio from 1 
can yield valuable 
information about $U_{e3}$ and $\cos \delta$. Deviations 
of $\alpha$
which are less extreme(between 0.7 and 1.5) can also probe 
very small pseudo-Dirac 
$\delta m^2$ (smaller than $10^{-12} eV^2$).
\end{itemize}  

Incidentally, in the last three cases, the results have 
absolutely no
dependence on the initial flavor mix, and so are 
completely free of any
dependence on the production model. So either one learns 
about the production
mechanism and the initial flavor mix, as in the first 
three cases, or one
learns only about the neutrino properties, as in the last 
three cases.
To summarize, the measurement of neutrino flavor mix at 
neutrino telescopes
is absolutely essential to uncover new and interesting 
physics of neutrinos. 
In any case, it should be evident that the construction of 
very large neutrino telescopes  is a ``no lose'' 
proposition.



\section{Acknowledgments}
This talk is based on work done in collaboration with
John Beacom, Nicole Bell, Dan Hooper, John Learned, Werner
Rodejohann and 
Tom Weiler. I thank them all for a most
enjoyable collaboration.
I would like to thank  the 
organizers of CTP 2009 for the
opportunity  to present this talk as well as their hospitality 
and for providing 
a most stimulating atmosphere during the meeting.
This work was supported in part by U.S.D.O.E. under grant 
DE-FG02-04ER41291.

\end{document}